# NAVIGATING THE CRYPTOCURRENCY LANDSCAPE: ISLAMIC PERSPECTIVE FOR 6TH INTERNATIONAL CONFERENCE ON ISLAM AND LIBERTY

Hina Binte Haq*[1], Syed Taha Ali[2]

## ABSTRACT

Almost a decade on from the launch of Bitcoin, cryptocurrencies continue to generate headlines and intense debate. What started as an underground experiment by a rag tag group of programmers armed with a Libertarian manifesto has now resulted in a thriving $230 billion ecosystem, with constant on-going innovation. Scholars and researchers alike are realizing that cryptocurrencies are far more than mere technical innovation; they represent a distinct and revolutionary new economic paradigm tending towards decentralization. Unfortunately, this bold new universe is little explored from the perspective of Islamic economics and finance. Our work aims to address these deficiencies. Our paper makes the following distinct contributions We significantly expand the discussion on whether cryptocurrencies qualify as 'money' from an Islamic perspective and we argue that this debate necessitates rethinking certain fundamental definitions. We conclude that the cryptocurrency phenomenon, with its radical new capabilities, may hold considerable opportunity which merits deeper investigation.

## INTRODUCTION

Bitcoin went live in January, 2009, the result of an underground experiment by a group of programmers with a Libertarian manifesto, and shortly reached parity with the US dollar in 2011.[1] It has since sparked countless scandals, spawned numerous imitators, and now dominates a thriving $230 billion cryptocurrencies ecosystem[2] with active ongoing technological and financial innovation.

Despite numerous obituaries in the mainstream economics press[3], Bitcoin continues to confound its critics. Currently Bitcoin trades at around the $6500 mark alongside other prominent cryptocurrencies such as Ethereum ($204), Dash ($154), Cash ($121), Monero ($104), Litecoin ($52), and Ripple ($0.46).[4] It is legal to use Bitcoin for payments and trading purposes in several countries, including the United States, Canada, the United Kingdom, Australia, Turkey, and Brazil, and it has notably attained legal tender status in Japan and Germany.[5]

More importantly however, even as it continues to make headlines, Bitcoin has generated intense debate in economics and policy circles. Here it is now widely understood that the Bitcoin is not a mere innovation in payment technology, but rather represents a distinct and revolutionary new economic paradigm. Bitcoin's remarkable innovation is best articulated by its pseudonymous creator, Satoshi Nakamoto, who introduced himself thus on the Cypherpunk mailing list almost ten years ago: *"I've been working on a new electronic cash system that's fully peer-to-peer, with no trusted third party."*[6]

With its marked shift towards decentralization, Bitcoin has forcefully reignited the century-old debate between the Keynesians and the Austrian regarding government intervention in the economy. Nakamoto himself states in an oft-quoted forum post, *"The root problem with conventional currency is all the trust that's required to make it work. The central bank must be trusted not to debase the currency, but the history of fiat currencies is full of breaches of that trust. Banks must be trusted to hold our money and transfer it electronically, but they lend it out in waves of credit bubbles with barely a fraction in reserve."*[7]

This worldview has numerous supporters, including former US Vice-President Al Gore who describes himself as *"a big fan of Bitcoin"* and echoes Liberal philosopher John Locke when he comments: *"Regulation of money supply needs to be de-politicized."*[8] Economics journalist Matthew Bishop compares Bitcoin's emergence to the recent "resurrection" of gold, and cites the two as *"a response to falling confidence in the soundness of government-backed 'fiat' money in an era of quantitative easing."*[9]

On the Keynesian end of the spectrum, economist Robert Shiller dismisses Bitcoin as a trend similar to the bimetallism fad of the 19th century; economist Nouriel Roubini is very vocal in criticizing the cryptocurrency universe as a bubble and *"the mother of all price manipulations"*.[10] Former chief economist at the IMF, Kenneth Rogoff, predicts that governments will eventually take over cryptocurrencies, much as they did in the case of standardized coinage and paper currency, both of which also originally started out as private sector innovations.[11]

Unfortunately, this bold new experiment in money is little explored from an Islamic economics perspective. Efforts in this domain have mostly been high level and non-technical discussions, focusing overwhelmingly on Bitcoin's role as a volatile commodity rather than a medium of exchange. Our work attempts to address this deficiency with a more wide-ranging and thorough treatment. Specifically, we significantly expand the discussion on whether cryptocurrencies qualify as 'money' from an Islamic perspective. Towards this end we revisit certain fundamental notions in the debate and highlight potential advantages of a 'Bitcoin standard' that may appeal to Islamic economics.

The rest of the paper is structured as follows: we start with a short primer on Islamic economics fundamentals and cryptocurrencies. Next we summarize existing research in this domain and categorize the various fatwas relating to cryptocurrencies. This is followed by a detailed consideration of recent arguments for and against cryptocurrencies as a legitimate Islamic currency.

We find that even as cryptocurrencies, by virtue of their radical design, present unique challenges to conventional systems, they also offer unprecedented opportunities. Specifically, from an Islamic economics perspective, cryptocurrencies resist the outright excesses of conventional fiat systems, the high costs and uncertainty of commodity money, and may therefore represent a useful and important step towards a credible Islamic system.

We hope our humble effort contributes to the ongoing debate in this exciting new domain.

## BACKGROUND

**A Primer on Islamic Economics**

The field of modern Islamic economics is widely considered to have emerged in the twentieth century, primarily due to the pioneering efforts of Iranian cleric Baqir Sadr and Pakistani theologian Abu- 'Ala Mawdudi.[12] This field is considered a sub-domain of Islamic jurisprudence (fiqh), which aims to harmonize economic thought and practices with Islamic ideals and behavioral norms as spelt out in Islamic scripture (Quran and Sunnah). Proponents of Islamic economics typically promote it as a legitimate third alternative to the mainstream systems of capitalism and socialism, as a middle way, which avoids the excesses of the two.[13]

We quote here from Askari, Iqbal, Mirakhor regarding the ideals of Islamic law: *"The overall aim or objective of Islamic law—that is, the concept of maqasid-al-Shariah—is to promote the welfare of humankind and prevent harm by preserving the faith, lives, intellect, prosperity, wealth, and interests of future generations. The preservation of these promotes society and its interests. The achievement of society's interests (maslahah) is essentially the same as maqasid; they are one and the same."*[14]

In light of these aims, Islamic scholars have proposed various guidelines for an ideal economy, including rules regarding property, framing of contracts, codes of conduct for investment and markets, workplace ethics, and wealth distribution, etc. There is a widespread consensus on certain key practices, sufficiently discussed and explained in the literature, such as the following:

- prohibition on usury (Riba) and its financial derivatives
- encouragement of participatory banking (i.e. profit and loss sharing)
- prohibition on gambling and speculative activities (futures, casinos, zero-sum games, traditional insurance, etc.)
- prohibition on earning from unethical and immodest activities (e.g. adult industry, music, etc.)
- prohibition on trading in forbidden goods (e.g. alcohol and pork)
- restricting monetary transactions to existing goods and services
- compulsory charity (Zakat) [15]

**A Primer On Cryptocurrencies**

Here we present a short non-technical overview of Bitcoin as a representative cryptocurrency, followed by a consideration of its security properties.

The Bitcoin payments system may be visualized as a distributed and highly synchronized ledger, which tracks the ownership of virtual currency units (or bitcoins). To transact in bitcoins, users create a pair of cryptographic credentials, a public and a private key. The public key is used to generate a Bitcoin address, which is the equivalent of a personal bank account number, while the private key enables users to spend the coins associated with an address. Bitcoin *transactions* are simply statements listing the Bitcoin addresses of the sending and receiving parties, the amount to be transferred. This entire statement is endorsed by the sender's private key, thereby authorising the transaction.

Users create and dispatch transactions on Bitcoin's global peer-to-peer network. This network is operated by voluntary participants (termed *peers*), running the Bitcoin protocol and maintaining a local copy of the Bitcoin blockchain. This is an authoritative ledger with a record of all prior transactions authorized by the network. Each peer independently verifies every incoming transaction against this ledger before accepting it as valid. Certain peers, referred to as *miners*, collect user transactions in batches, or *blocks*, and then periodically compete in a decentralized lottery protocol to insert their block in the blockchain. Bitcoin incentivizes miners by rewarding the winning parties with newly created bitcoins and awarding them any transaction fees accumulated from their blocks.

This mechanism enables independent and mutually untrusting parties to transact fairly without reliance on a trusted third-party. If any party tries to re-spend money from an address that has already been spent, nodes in the network will detect the discrepancy during the transaction verification process. Furthermore, any node in the network can participate in the mining protocol to select which transactions are inserted into the canonical record. This strategy prevents a single point of failure in the network and defeats censorship.

The protocol relies for its guarantees on network-wide consensus. As long as a majority of the nodes in the network are honest, malicious parties are unable to double-spend funds, censor or reverse transactions, or generate extra new currency units. Bitcoin also affords users a degree of anonymity in that Bitcoin transactions do not require any personal information regarding the sending and receiving parties.

However, certain weaknesses in the protocol have been discovered recently: for instance, anonymity in Bitcoin can be compromised by employing clustering and forensics techniques[16]. Second, certain parties can accumulate excess computing power and thereby exercise a disproportionate influence over the mining process. The mining process is also compute-intensive and, with Bitcoin's soaring popularity, has reached unsustainable levels of energy consumption [17].

**Prior Work**

Here we briefly summarize how various scholars and researchers have assessed cryptocurrencies from an Islamic perspective.

**Research Literature**

Existing work in this domain has mostly adopted a piecemeal approach in that researchers have focused on select aspects of cryptocurrencies to the exclusion of others. Their opinions have mostly been favourable and several have suggested that concerns regarding the permissibility of cryptocurrencies may be resolved with technical modification and policy regulation[18].

Meera[19] discusses the requirements for money in Islam and concludes that cryptocurrencies fall short on account of having no intrinsic value of their own. However, discussions by Evans[20], Muedini[21] Oziev[22] argue that Bitcoin (or a similar system), with a few alterations, might actually prove a more appropriate medium of exchange from an Islamic perspective than existing interest-backed fiat currency. A pioneering effort in this regard is due to Bergstra[23], who references Maulana

Mawdudi's discourse and recasts Bitcoin as a money-like commodity which does not intrinsically support debt-creation and is therefore not fundamentally opposed to the Islamic vision.

Abu Bakr[24] argues that aspects of uncertainty (gharar) and risk within Bitcoin due to its price volatility, most specifically its transaction mechanics, pose challenges from an Islamic perspective. Similarly, Bergstra also notes that the random lottery-like mechanism behind mining may be problematic in that it is reminiscent of gambling, and he suggests that the mining function be expressly relegated to trusted parties.

Abdullah[25], Zubaidi[26] and Yousaf[27], propose that the Islamic states co-opt cryptocurrency technology to create national currencies that are alternatives to existing fiat currencies. They propose numerous modifications to decentralize cryptocurrencies, including reviving the role of central banks and authoritative entities within the ecosystem, or backing currency units with commodities like gold or silver.

**Legal Opinions**

A number of scholars have issued legal opinions (fatwas) regarding Bitcoin. These efforts also focus on specific aspects of cryptocurrencies.

Mufti Taha Karaan and Mufti Siraj Desari (South Africa), Darul Ifta, DUZ (South Africa), Professor Monzer Kahf argue that owning and trading Bitcoin as permissible (mubah), based on social concurrence (istilah) and common usage (ta'amul), however they are not yet convinced about its standing as a currency.[28]

Dar ul Ihsan (South Africa) recommends Muslims abstain from using Bitcoin until more clarity is achieved regarding cryptocurrencies. They base their fatwa on the suspicion that, given their lack of intrinsic value, cryptocurrencies may prove to be a pyramid scheme. Sheikh A. S. Abu Ghuddah (Syria) also currently deems Bitcoin impermissible and un-Islamic due to excessive uncertainty regarding its value and the associated security risks, but he does not prohibit it. Diyanet (Turkey) similarly proscribes Bitcoin and also cites its widespread use in criminal activity.[28]

Wifaqul Ulama, Sheikh Haitham al-Haddad and others from UK, Darul Iftaa, (India), along with Sheikh Imran Husain (Trinidad), Mufti Shawki Alam and Magdi Ashour (Egypt), Assim al-Hakeem (Saudi Arabia) have declared Bitcoin as forbidden (haram) because it is not issued by the state, the mining process is similar to gambling, excessive price volatility is a grave risk, and that Bitcoin is used to fund terrorists.[28]

**THE COMMODITY DEBATE**

**Money in Islam**

Money has taken various forms over history, ranging from barter (the exchange of goods), to commodities (e.g. grains, dates, etc.), precious metals (gold, silver), standardized coinage, gold-backed paper currencies, and now, fiat systems. Recent years have witnessed renewed interest in the question of what constitutes money from an Islamic perspective. In an informative survey on this question, Haneef and Emad categorize scholarly views on this question into two main camps: those who limit money only to gold and silver and those who do not.[29] We briefly consider their arguments here and present a case for cryptocurrencies.

Proponents for a definition of money that is exclusive to gold and silver cite numerous Quranic verses, which equate money with gold and silver (e.g. al-Tawbah: 34, Al-Imran: 75, 91, and Yusuf: 20). Moreover, the calculation of zakat on wealth is based on gold and silver, as is that of blood-money (diyah). The Prophet himself as well as the guided caliphs are also known to have explicitly approved the use of gold and silver as money. Furthermore, gold and silver possess intrinsic value and are known for their ability to preserve their value over long periods of time.

On the other hand, in early Islamic history we find various incidents in support of a more flexible definition for money. At the time of the Prophet (PBUH), Muslims were known to use gold, silver, and Byzantine coins. Limited amounts of copper coins (fulus) were also in circulation. There is also anecdotal evidence from that period whereby remuneration was paid using rice, dates, wheat, barley, and salt. In one instance, Hazrat Umar, during his caliphate, discussed the idea of using leather from camels as money but the Sahabah (Companions) disagreed. This example is revealing, because the disagreement was not on the grounds that the idea was not permissible, but because implementation of it would lead to a shortage of camels.

The introduction of fulus was necessitated in certain cases due to the difficulties of relying solely on gold and silver as money in instances such as shortages. Ibn Taymiyyah (d. 728 A.H) accepted fulus as money: *"The authority should mint the 'fulus' coins (other than gold and silver) according to the just value of people's transactions, without any injustice to them"*. However it is pertinent to note that fulus circulated in limited amounts in Ibn Taymiyyah's lifetime and was not the primary currency in this period.[30]

However, fulus created controversies of its own. During the reign of Muhammad b. Ali (806 AH), large quantities of fulus in circulation led to hyperinflation. This was accurately diagnosed by the historian Al-Maqrizi, and likely shaped his opinion that gold and silver should be the primary forms of money, whereas small quantities of fulus in circulation is acceptable[31]

Imam Ghazali and Ibn Taymiyyah also condemned the issue of debasement by alloying gold or silver with an inferior metal to mint coins (maghshzish money) or by reducing their weight. However, Imam Ghazali holds the view that such coins are acceptable if issued by a central authority[32].

Scholars who subscribe to the view that money can be apart from gold and silver cite the principle of ibahah, according to which a matter is considered impermissible only if there is a prohibition to that effect in the Quran or Sunnah. These scholars also maintain that such issues fall under al-masalih al-mursalah, i.e. it is to the discretion of the authorities of the time to decide on these matters in the interest of social welfare. Furthermore, Islam aims to promote harmony in society and ease hardship, a goal that might be hindered, especially today, due to the limited supply of gold and silver in the world. However, an important consideration in the case of a new currency is that other Islamic injunctions such as Zakat and the prohibition on usury still need to be strictly observed.[31]

**Bitcoin and gold**

If we turn to cryptocurrencies today, particularly Bitcoin, we observe that Bitcoin meets most contemporary notions of a currency: it serves as a medium of exchange (as an intermediary in trade), a unit of account (it is quantifiable), it is divisible into smaller units (satoshis), it is fungible (a bitcoin is mutually interchangeable with other bitcoins), it is durable, and it is portable. The only perceived negative is that Bitcoin's price volatility compromises its role as a store of value. However, proponents argue that as cryptocurrencies become more established, we can expect to see prices stabilize in the long term. [33]

This is also widespread discussion in the mainstream economic press regarding Bitcoin's affinity with gold, due primarily to the fact that Bitcoin has emerged as a competitor to the established fiat paradigm. Satoshi Nakamoto, himself, articulates the metal-like nature of Bitcoin in responding to a forum post as follows:

*"As a thought experiment, imagine there was a base metal as scarce as gold but with the following properties:*

*boring grey in colour*
*not a good conductor of electricity*
*not particularly strong, but not ductile or easily malleable either*
*not useful for any practical or ornamental purpose*
*and one special, magical property:*
*can be transported over a communications channel*

*If it somehow acquired any value at all for whatever reason, then anyone wanting to transfer wealth over a long distance could buy some, transmit it, and have the recipient sell it.*

*Maybe it could get an initial value circularly as you've suggested, by people foreseeing its potential usefulness for exchange (I would definitely want some). Maybe collectors, any random reason could spark it."*[34]

Indeed, Bitcoin shares several properties with gold that appeal to the Libertarian worldview. Both Bitcoin and gold are stateless. They incur considerable mining costs, and they are scarce, i.e. they cannot be created "out of thin air" by central banks as in the case of fiat. Due to this last property, countries on a gold standard have had a considerably limited scope of monetary policy as compared to the case of fiat.

However, there are significant differences nevertheless. Whereas gold remains a physical commodity, Bitcoin is digital and intangible. Gold has had an intrinsic value over the centuries, whereas there is debate over how to quantify or even define such a value for Bitcoin. However, the most striking dissimilarity is their supply and demand mechanisms. While gold is mined around the world and its supply increases in an unpredictable manner, bitcoins are mined in fixed amounts as specified in the software with a predetermined future total of 21 million units.[35]

**Is Bitcoin a commodity?**

Commentators have further explored Bitcoin's resemblance to gold and argue that criticisms of Bitcoin may not be taking full account of its strengths. For instance, the argument that Bitcoin lacks intrinsic value is plausible at face value, but relies on a narrow conception of 'value'. Bitcoin possesses inherent value, as Nakamoto argued, in that it can be transferred online securely, efficiently, and rapidly at very low cost. Bitcoin also reduces reliance (and consequently the associated costs) of trusted third parties and central banks which also translates to value.

Likewise, it may be argued that even though bitcoins are a purely digital construct, significant real world cost and effort is expended in creating new units of the currency, a process that some argue is not unlike that of mining physical metals. Miners globally compete in solving cryptographic puzzles to add blocks to the blockchain and this requires immense computational processing. This further entails significant costs in terms of hardware, infrastructure, and electricity. It is estimated that, at the time of this writing, mining has an annual electricity consumption of 73.12 TWh (almost equivalent to annual consumption of countries like Austria and Philippines).[36]

Moreover, much like commodities, the ownership of bitcoins can be transferred, bitcoins can be consumed (by spending), and also destroyed (a process known as *burning* whereby bitcoins are transmitted to an unspendable address).[37]

Some scholars have ventured even further to suggest that cryptocurrencies, far from being simply 'digital' or 'virtual' money may actually represent a fundamentally new paradigm, which advances our historical understanding of money.[38]

In one of the earliest technical studies on this topic, Bergstra classifies Bitcoin as a **money-like informational commodity**[39], with particular relevance to an Islamic economic system[23]. Such a commodity has the following peculiar properties: first, it has an *"exclusively informational status."* Second, the "*moneyness*" of such a commodity *"correlates with [its] usage and acceptance by a significant and relevant fraction of the public."* Finally *"ownership"* of such a commodity is *"identical to control [of] (or access) to that quantity."* This last point has particular relevance from an Islamic perspective since it is technically *"pointless"* to loan an informational commodity because "*all transfers of access have the same status."* In short, money-like informational commodities do not intrinsically lend themselves to borrowing or debt as in the case of fiat or fiduciary money.

In a similar vein, Selgin revisits the fiat vs. commodity debate and suggests that this conventional dichotomy is false. He focuses on two key properties of money, scarcity and non-monetary use. Bitcoin is scarce in a similar sense to commodity money by virtue of its design, whereas fiat money relies on a contrived scarcity, one enforced by the state. However, unlike commodity and like fiat, Bitcoin does not possess any non-monetary application. Selgin argues that such an item qualifies as a legitimate third type of money, a **synthetic commodity money**[40] since *"it shares features with both commodity money and fiat money, as these are usually defined, without fitting the conventional definition of either"*. A similar example of such a money is the Iraqi Swiss dinar, which circulated in northern Iraq for over a decade in limited supply and without any government backing.

Selgin suggests that synthetic commodity money holds significant advantages over commodity and fiat money alike. In the case of commodity money like gold, expanding the money supply entails significant real cost (an issue that Friedman considered the *"fundamental defect"* of the gold standard[41]). Second, gold is vulnerable to shortages and supply-side shocks. Cryptocurrencies can be carefully designed to address both these issues.

The difference with fiat is also stark. Synthetic commodity money is not fatally reliant on state intervention and exempt from monetary experiments undertaken by central banks. As per Selgin, this *"supplies the basis for what Buchanan calls an "automatic" as opposed to a deliberately "managed" monetary system"* where monetary policy is confined to operate within strictly defined limits. Furthermore, Selgin maintains: "*free banking and an inelastic synthetic commodity standard is therefore capable in principle of automatically promoting macroeconomic stability.*"[40]

The point regarding the limitations on central banks is further explored by Weber[42] who undertakes an interesting experiment: what would happen if the world were to abandon the existing fiat regime in favour of a 'Bitcoin standard'? Given Bitcoin's affinity with gold, Weber takes the classical gold standard era in the West as the starting point of his investigation, specifically the years 1880-1913. In such a situation, Weber suggests, much as with gold earlier, there would be select media of exchange: first, Bitcoin itself and second, fiduciary currencies issued by central banks and regular banks.

Whereas, the utility of Bitcoin is such that it does not necessitate the creation of fiduciary currencies (indeed Bitcoin's original philosophy is antithetical to the idea), Weber argues that monetary authorities may rely on Bitcoin-back fiduciary currencies to finance fiscal deficits. And it is on this point that we observe a key difference between the gold standard and the hypothetical Bitcoin standard.

During the historical gold standard period, a central bank could set its interest rate differently from that of other countries to influence its domestic economy. Normally gold would flow to the country where it would fetch the highest rate of return, but this flow across borders was curbed by the high costs of gold arbitrage (costs due to shipping, insurance, etc.). This effectively provided central banks some flexibility to manipulate their interest rates.

However, in the case of Bitcoin arbitrage costs would be trivial. This would defeat any effort to set interest rates different from those of other countries, and spot exchange rates for all fiduciary currencies would be one-to-one. This reality also severely restricts the ability of a central bank to act as lender of a last resort in case of bank runs or financial crises.

This essential limitation on debt creation afforded by Bitcoin is again in stark contrast to the case of fiat where central banks can dramatically expand the money supply and create new debt at will to influence their economies. This argument may appeal to proponents of Islamic economics, particularly those who argue for a return to a gold standard because it does not permit arbitrary debt creation.

**CONCLUSION**

We hope our discussion thus far provides the reader with an appreciation for the diverse research being done in this domain and a sense of the radical potential of cryptocurrencies.

We have focussed our discussion primarily on the question of whether cryptocurrencies qualify as 'money' from an Islamic perspective. To this end we have examined the historical debate within Islam where we observe considerable flexibility in choice of currency. We have also explored the close similarities and differences between Bitcoin and gold, and then delved into the emerging debate which posits Bitcoin as a wholly new type of money with unique properties of its own and compelling advantages over both commodity and fiat money. We also briefly explore the ramifications of a Bitcoin standard.

We believe this debate about the 'moneyness' of cryptocurrencies is central to its utility to Islamic economics. If this debate is satisfactorily resolved, we believe remaining problematic issues (such as the element of gambling, the ecological footprint of mining) may be addressed with the help of new technical developments and policy regulations.

We hope our effort contributes to greater dialogue and discussion in this domain.

**ACKNOWLEDGEMENT**

The authors would like to thank Mr. Ali Salman for his immense assistance in providing source material and directions for this effort.

# AUTHORS

Author* Hina Binte Haq, PhD Student, School of Electrical Engineering and Computer Sciences, NUST, H-12, 44000, Islamabad-Pakistan. E-mail: hhaq.dphd18seecs@seecs.edu.pk

Author[2] Syed Taha Ali, Assistant Professor, Department of Electrical Engineering, School of Electrical Engineering and Computer Sciences, NUST, H-12, 44000, Islamabad-Pakistan. E-mail: taha.ali@seecs.edu.pk